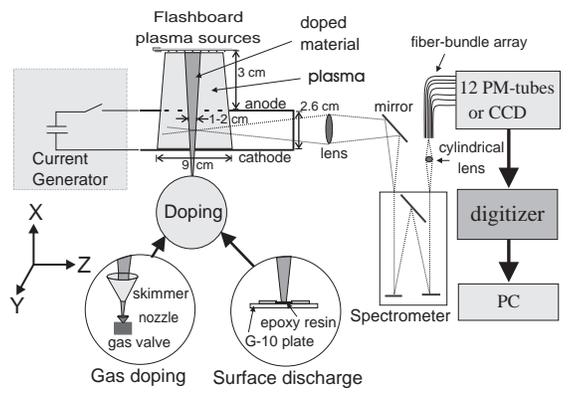

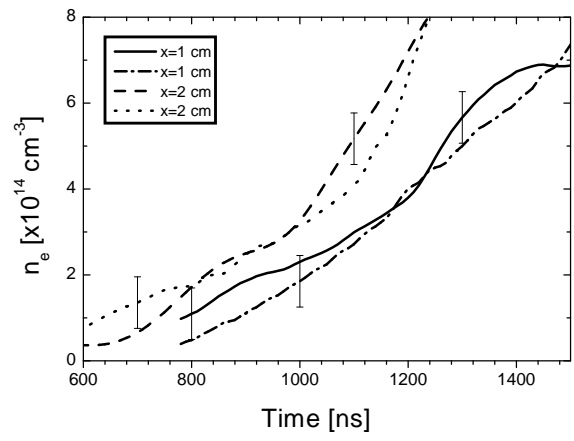

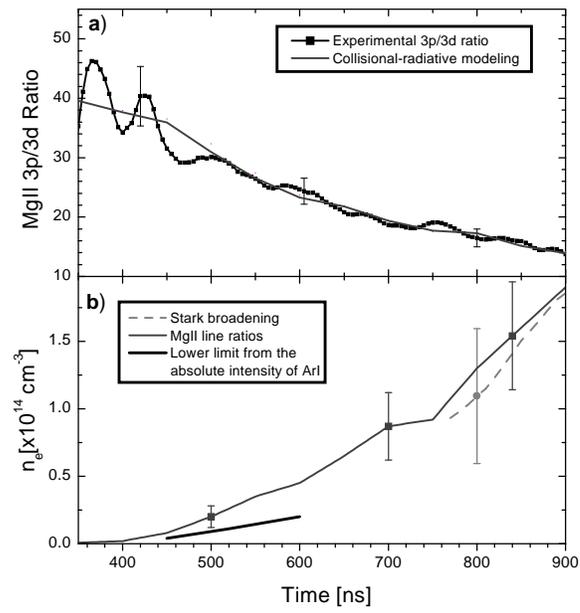

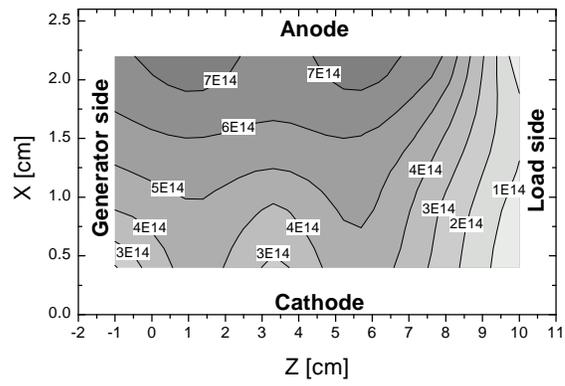

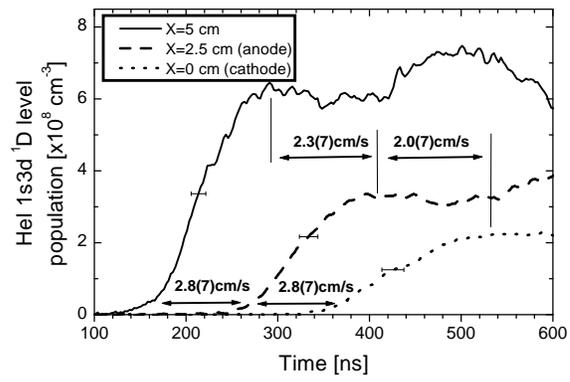

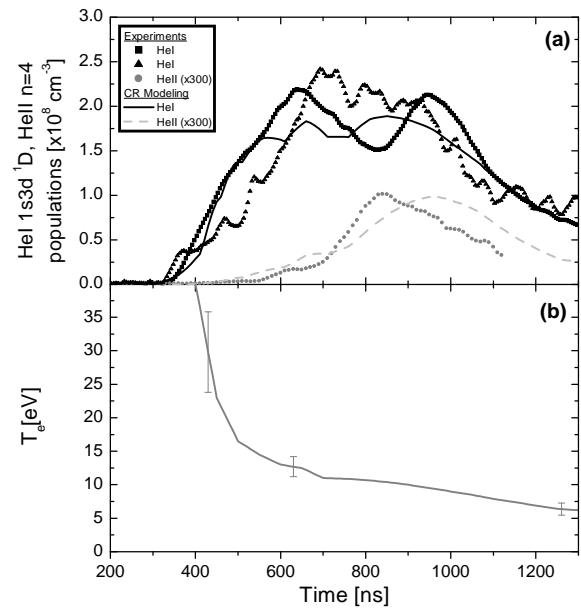

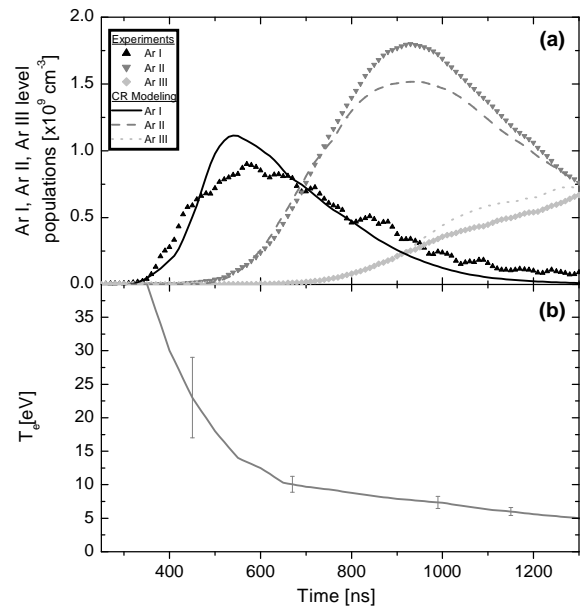

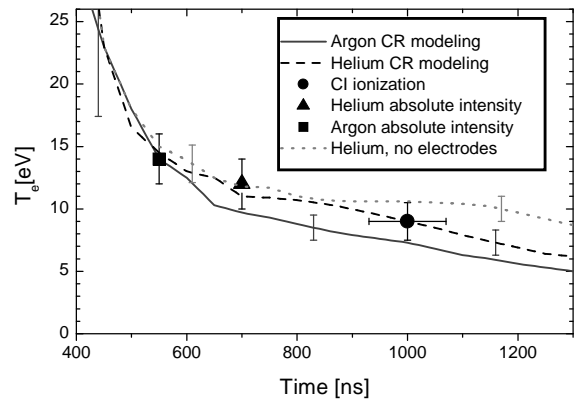

# Spectroscopic investigations of a dielectric-surface-discharge plasma source

R. Arad[*], K. Tsigutkin, Yu.V. Ralchenko, and Y. Maron

*Faculty of Physics, Weizmann Institute of Science, 76100 Rehovot, Israel*

Spectroscopic investigations of the properties of a plasma produced by a flashboard plasma source, commonly used in pulsed plasma experiments, are presented. The plasma is used to prefill a planar 0.4-µs-conduction time plasma opening switch (POS). A novel gas-doping technique and a secondary surface flashover plasma source are used to locally dope the plasma with gaseous and solid materials, respectively, allowing for spatially-resolved measurements. The electron density, temperature, and plasma composition are determined from spectral line intensities and line profiles. Detailed collisional-radiative modeling is used to analyze the observed line intensities. The propagation velocity and divergence angle of various ions are determined from time-of-flight measurements and Doppler broadening of spectral lines, respectively. This allows for distinguishing the secondary plasma ejected from the POS electrodes from the plasma of the flashboard source.

PACS numbers: 52.50.Dg, 52.70.-m, 52.70.Kz, 52.75.Kq

[*]Electronic mail: fnarad@plasma-gate.weizmann.ac.il

## I. Introduction

The need for reproducible plasma sources with various properties (electron density, electron temperature, plasma composition and uniformity) for different pulsed-power applications such as ion diodes [1] and plasma opening switches (POS) [2] encouraged the development of a variety of plasma sources. Among the most popular schemes are plasma guns [3], flashboards [4,5], gaseous plasma sources [6,7], and explosive emission sources [8]. In plasma (or cable) guns, a flashover occurs across the insulator at the end of a high-voltage cable. Flashboards are printed circuit boards consisting of a number of chains, each with a set of flashover gaps in series. A high-voltage pulse is applied to the ends of each chain and a surface breakdown is initiated across the gaps.

The primary goal of this work is to study the plasma properties of a flashboard plasma source as a first step towards a thorough investigation of the physics of the interaction between plasmas and magnetic fields in general, and in particular that of POS's. In our setup the flashboard plasma source is used to prefill a planar POS (0.4-µs-conduction time and 160 kA peak current) coupled to an inductive load.

Previous experimental studies of flashboards include a detailed investigation of the plasma flow velocity and uniformity for various driving circuits and flashboard configurations [5] using electrical probes. A recent experiment [9], in which the electron density was determined both by spectroscopic measurements and by electrical probes, showed that electrical probes underestimate the electron density of a flashboard plasma by up to an order of magnitude. A spectroscopic investigation of the electron density, temperature, and plasma composition was carried out on a flashboard driven by a relatively low current through the discharge chains [10]. In that work, the peak electron density $n_e$ was determined to be $2\times10^{13}$ cm$^{-3}$ and the plasma was found to mainly consist of CII and CIII. However, it should be noted that the density of CIV and CV could not be determined in those experiments due to the low electron temperature. Plasma guns being closely related to flashboards have been studied [3,10,11,12] using various techniques that showed the plasma mainly consists of carbon ions with electron densities up to $\approx 10^{14}$ cm$^{-3}$. Interferometry was implemented to determine the line-integrated electron density and its spatial distribution in a few POS experiments [13,14,15] where either plasma guns or flashboards were used. As will be shown below, determining the electron density, temperature and plasma composition in such plasma sources is a complex problem involving various processes such as plasma flow, ionization, and secondary plasma creation. To the best of our knowledge, these characteristics have as yet not been determined satisfactorily in flashboard and plasma gun plasmas. In this work we determine these plasma properties using different spectroscopic observations together with atomic-physics modeling.

High resolution spectroscopy is employed to measure line profiles and intensities of a variety of spectral lines. Spatial resolution along the line of sight is obtained by doping the plasma with different elements and observing the characteristic emission from the doped species. To this end, we employ our newly developed gas injection [16] and surface flashover techniques to locally dope the plasma with gaseous and ionic material, respectively. The versatility of the doping methods also allows us to perform measurements using a wide variety of dopant elements which enables us to study many important plasma parameters. Furthermore, by measuring and analyzing spectral lines belonging to a few different elements we are able to reduce the errors introduced by the uncertainty of the different atomic rates.

The plasma electron density is obtained from Stark broadening of the $H_\alpha$ and $H_\beta$ lines of hydrogen. A second complementary method, based on spectral line ratios of MgII, is used to study the electron density at early times when $n_e < 10^{14}$ cm$^{-3}$ and the Stark broadening yields less reliable results. It was verified that for high electron temperatures ($T_e > 10$ eV) and low electron densities, the ratio between the MgII 3p and 3d level populations depends primarily on $n_e$. A third method compares the absolute intensity of doped ArI lines with collisional-radiative calculations and the known argon density. The use of a few methods provides us with a more reliable electron density and enables measurements in a wide range of densities ($1 \times 10^{13}$ -$2 \times 10^{15}$ cm$^{-3}$).

The electron temperature is obtained from the temporal evolution and absolute intensities of spectral line emissions of various doped gases. The observed evolution of spectral lines from different charge-state ions is compared to collisional-radiative (CR) calculations [17]. Other, more conventional methods for determining the electron temperature such as spectral line ratios, could not be fully implemented here. A typical procedure would be to compare the line intensities from levels with significantly different energies both of which should not be too high. Here, this requirement could not be fulfilled due to the rapid ionization of low charge-state species and the possible presence of non-thermal electrons.

The plasma composition is obtained from the absolute intensities of spectral lines belonging to a variety of species that make up the plasma. To this end, CR calculations that are based on the determined electron density and temperature are used for calculating the ratio between the excited level populations and the total density of the specific ion. The effects of charge exchange between CV and hydrogen are examined and found to significantly affect some of the CIV excited level populations, allowing the CV density to be determined. It is shown that the plasma mainly consists of protons, hydrogen, and carbon ions. By comparing the plasma composition with and without the POS electrodes in place we are able to determine which ions originate from the flashboard and which are coming off the POS electrodes. The use of the gas doping also allows for studying the non-radiating proton plasma at the front of the expanding plasma cloud by observing the excitation of the doped gas by the electrons in the proton plasma. We believe that the diagnostic methods described here can be used for studying plasma sources and transient plasmas in general.

## II. Experimental Setup

Two flashboards are used to generate the plasma. A single 2.8-µF capacitor charged to 35 kV drives the flashboards via sixteen 75-Ω cables, giving a peak current of 6 kA per chain at t = 1.2 µs. The flashboards are positioned next to each other in such a way that the peak plasma density is generated in the middle of the plasma volume. Each of the flashboards has eight chains and each chain consists of 8 copper islands on a G-10 insulator of a 0.07-cm thickness with a 1.5-cm gap between the chains. The islands are 0.5 cm in diameter and are separated by 0.05-cm gaps (in our experiments the flashboards are not coated with carbon spray as is commonly done).

The flashboards are placed 3 cm above the 8-cm-long region of the POS transparent anode, see Figure 1. The planar POS electrodes are 14 cm wide (along the y-direction) and are 2.6 cm apart. In the POS region, the electrodes are highly transparent and consist of 0.1 cm-diameter wires separated by 1.4 cm. This high transparency ensures free plasma flow through the anode and little plasma stagnation near the cathode. The flashboard is operated ≈1.2 µs before the high-current pulse of the POS is applied. The following coordinates are defined: $x = 0$ is the cathode, $y = 0$ is the center of the electrodes, and $z = 0$ is the generator-side edge of the anode transparent region.

The gas doping arrangement, consisting of a fast gas valve, a nozzle and a skimmer, is mounted below the cathode on a moveable stand that allows for 2D movement. The dopant-gas beam was diagnosed using a specially-designed, high spatial-resolution ionization probe array [16] that allows the absolute gas density and the beam width to be determined. The gas density can be varied from approximately $10^{13}$ to $3 \times 10^{14}$ cm$^{-3}$ by changing the time delay between the operation of the gas valve and the flashboard. The full width half maximum (FWHM) of the gas beam perpendicular to its injection could be varied from 1 to 2 cm by changing the aperture of the skimmer.

Alternatively, for doping solid materials such as magnesium, we used an electrical discharge over an epoxy resin mixed with the desired element. A small G-10 board was placed 2 cm below the cathode and a discharge was driven by a 2 µF capacitor charged to 6 kV. The hereby-produced plasma was diagnosed spectroscopically for various time delays between this discharge and the operation of the flashboard. An upper limit of the electron density of the doped material is determined from the absolute intensity of the MgII 3p-3s transition without the operation of the flashboard plasma. The emission of this line is found to be very weak. To assure that the weak emissions do not result from an exceedingly low electron temperature (<1 eV) with a relatively high electron density, we compare the emission with and without the application of the flashboard and make use of the fact that once the doped column mixes with the flashboard plasma the electron temperature of the combined plasma is at least 3 eV and the combined electron density is ≈$10^{13}$ cm$^{-3}$ at t=450±30 ns. Then, by performing measurements with different time delays between the doping discharge and the flashboard operation we obtain an upper bound on the electron density of the doping discharge as a function of time. For the short time delays used in the present experiments, the electron density of the plasma created by the doping discharge was found to be less than $1 \times 10^{13}$ cm$^{-3}$, ensuring no significant effect on the flashboard plasma parameters.

For the short time delays used here, the neutral and electron densities of the doped column are small enough inside the A-K gap to assure they do not significantly perturb the flashboard plasma. The density of neutrals such as MgI and CI are negligible in the A-K gap due to the rapid ionization caused by the flashboard-plasma electrons. The density of neutral hydrogen produced by the solid doping discharge is significantly lower than the hydrogen density ejected off the POS electrodes. We did not measure the density outside the A-K gap, however, based on comparison of measurements with different time delays we believe both the electron and the neutral densities to be <$10^{14}$ cm$^{-3}$ even 1 cm beyond the cathode.

Mirrors and lenses are used to direct the light onto a 1-meter spectrometer equipped with a 2400 grooves/mm grating. Observations along all three lines of sight are possible, however, in this work we present only observations performed along the y-direction. Observations along the x-direction are performed through the flashboards. A cylindrical lens images the output of the spectrometer onto a rectangular optical-fiber array at different magnifications allowing for observations with different spectral dispersions. The optical-fiber array consists of 12 vertical fiber stacks that deliver the light to 12 photomultipliers whose temporal resolution is 7 ns. The spectral resolution of our optical system is 0.07 Å.

The absolute sensitivity of the optical system was measured using a few absolutely-calibrated lamps. In order to reduce the error in the relative calibration at different wavelengths we measured the light intensities of different spectral lines that originate from the same excited level. We observed the CII 2837-Å and 6578-Å lines, the NII 6284-Å and 3919-Å, and the SiII 2072-Å and 4130-Å lines. An accuracy of ±10% for the relative sensitivity and ±30% for the absolute calibration was determined.

## III. Experimental Results

### A. Measurements and Data analysis

We observe the time-dependent intensities and spectral profiles of lines from the plasma constituents (mainly hydrogen, carbon, and oxygen) and the doped materials (such as helium, argon, and magnesium), as given in Table I. The line profiles are fitted using a Gaussian, from which the instrumental broadening is deconvolved. The width of the various lines is analyzed assuming Stark and Doppler broadening (all spectral lines used in this study are optically thin). The measured line intensities yield populations of the transition upper levels by using the Einstein coefficients of the transitions, the observed light emitting volume and the absolute sensitivity of the optical system. The level populations are then compared to time-dependent CR calculations in order to determine the plasma parameters such as $n_e$ and $T_e$. The level populations given throughout this paper are divided by statistical weight.

The typical velocities of species doped using the gaseous doping are $1 \times 10^5$ cm/s so that within 1 μs their motion is negligible. However, for the plasma and solid doping constituent the ion velocities are a few times $10^6$ cm/s so that their motion during the time of interest is significant. To study the ion fluxes we used the fact that the ratio between different level populations and the total ionic density equilibrate within times determined by the radiative decay of the level (typically < 10-20 ns). Thus, the observed evolution of the level populations reflects the variations of the ionic density, $n_e$ and $T_e$. Since (as will be shown bellow) $n_e$ and $T_e$ are determined from line widths, intensity ratios and line intensities of gaseous elements (whose motion is negligible), we could use the evolution of the observed intensities to determine the fluxes of the ions whose motion is significant.

In the CR modeling [17] the time-dependent atomic/ionic level populations are calculated from the following system of rate equations:

$$\frac{dN(t)}{dt} = A(t)N(t), \qquad (1)$$

where N(t) is the vector of all the level populations and A(t) is the rate matrix. The treatment is time-dependent (the rate equation system (1) is integrated explicitly), as is required for such highly transient plasmas. The number of atomic levels used in the calculations varies from about 10 for Hydrogen up to about 200 for the oxygen ions (OI-IX). The matrix A(t) depends on time via the electron density $n_e(t)$ and the time-dependent EED function. The basic atomic processes taken into account in the rate matrix are electron impact excitation and deexcitation, electron impact ionization and 3-body recombination, spontaneous radiative decays, radiative and dielectronic recombination, and charge exchange. It must be emphasized that in the present modeling we use the energy-dependent cross sections rather than Maxwellian-averaged rate coefficients, which allows for CR calculations with an *arbitrary* EED function. This is essential for studying the effects of the non-Maxwellian EED in the flashboard plasma.

The atomic data required for our collisional-radiative modeling are either taken from existing atomic databases or newly calculated using available computational tools. The National Institute of Standards Atomic Spectra Database [18] is the principal source of energy levels and oscillator strengths; the missing values are calculated with Cowan's Hartree-Fock relativistic atomic structure package [19] including configuration interaction and intermediate coupling. Since the available data on the electron impact excitation are rather limited, most cross sections are calculated with the Coulomb-Born-exchange code ATOM [20]. The excitation cross sections are mainly calculated in LS-coupling although, when necessary, other types of couplings are utilized (for example, jK-coupling for ArI). To improve the accuracy of dipole-allowed excitation cross sections, they are rescaled by the ratio between a more accurate oscillator strength, calculated by other methods (e.g., multi-configuration Hartree-Fock), and that calculated by ATOM (corrections are typically <50%). For some transitions between the levels with close energies this correction exceeded a factor of 2 (which may result in lower accuracy); however, the absolute magnitude of the corresponding cross sections is less important due to fast equilibration of the relevant level populations. Comparison between the rescaled excitation cross sections obtained with ATOM and available experimental data shows agreement within 20-30% for ions and within 50% for neutral atoms. For H and HeI-II more accurate excitation cross sections, calculated with the Convergent Close-Coupling method [21], are used to obtain the highest possible reliability. The use of accurate cross sections is especially important when the electron temperature is much smaller than the excitation threshold, and therefore the corresponding rate is mainly determined by the near-threshold cross section. The dipole-allowed cross sections for strong *two-electron* transitions are computed in the Van Regemorter approximation. The recommended ionization cross sections for the ion ground states [22] are also used for ionization from the excited states, the classical scaling $\sigma_{ion} \sim I^{-2}$ being implemented for the latter (I is the ionization energy). The Milne relation [19] is used to produce the radiative recombination cross sections from the Opacity Project results [23] for photoionization. The dielectronic recombination rates of Hahn [24] are used for most ions while for neutral and low-charge ions of argon the Burgess-Merts-Cowan-Magee parameterization [19] is utilized. Finally, the cross sections for the inverse processes are obtained using the principle of detailed balance which, although formulated for equilibrium plasmas, establishes relations between purely atomic quantities, and thus these relations are independent on particular plasma conditions such as, for example, the electron energy distribution function.

## B. Electron density

Stark broadening measurements of the $H_\alpha$ and $H_\beta$ profiles are analyzed self-consistently to give $n_e$ and the hydrogen velocity distribution along the line of sight. In this analysis, the Stark broadening is assumed to mainly result from the quasistatic Holtzmark distribution of the ionic microfields with a small correction due to the electron dynamics, treated under the impact approximation. The line width calculations are performed according to Ref. [25]. The effects of ion dynamics (negligible for $H_\beta$) are unimportant for $H_\alpha$ in this experiment due to the dominance of the Doppler broadening. The Doppler broadening results from the combined effect of the thermal velocity and the directed velocity, integrated along the line of sight. The $H_\alpha$ profiles are fitted by a Gaussian which is in good agreement with the observed profiles. Because of the large Doppler width of the hydrogen lines the use of Stark broadening for the determination of the electron density is limited to $n_e > 10^{14}$ cm$^{-3}$. The Stark broadening dependence on the effective charge of the ions in the plasma is accounted for, based on the determined plasma composition (see Sec. III.E).

The electron density measurements are performed in the POS anode-cathode (A-K) gap which is located 3-5.5 cm from the flashboard surface. $H_2$ and $CH_4$ doping is used to determine $n_e$ locally in 3D by observing the line emission from hydrogen produced from the dissociation of these molecules. Due to the complex chemistry of the doped $CH_4$ molecules, whose detailed modeling is beyond the scope of this work, we performed the measurements using different densities of $CH_4$ and $H_2$ doping and compared them to experiments without doping. We were careful to operate with gas densities that are significantly smaller than the original plasma density to minimize the effects of the unknown plasma chemistry.

The rise in the plasma electron density due to the ionization of these doped gases is estimated to be less than $5 \times 10^{13}$ cm$^{-3}$ due to the low density of the injected gas ($n_{gas} < 3 \times 10^{13}$ cm$^{-3}$). However, at t > 1 μs the density of hydrogen ejected from the POS electrodes becomes comparable to that of the doped hydrogen, resulting in line-integrated (along the y-direction) electron density determination.

The evolution of $n_e$ at z = 3.5, y = 0, and x = 1 and 2 cm is shown in figure 2. In these experiments $CH_4$ is injected into the entire POS region by removing the skimmer. The spatial resolution in these measurements is ±0.2 cm along the x-direction, ±0.5 cm along the z-direction, and spatially averaged along the y-direction. At each position two traces from two different discharges are shown to indicate the reproducibility $n_e$, found to be within ±10% at t > 1 μs.

The ratio between the MgII 3p and 3d level populations, determined from the intensities of the 3s-3p and 3p-3d transitions, is used in order to study $n_e$ at times earlier than 800 ns (prior to the ionization of the MgII). This method relies on the fact that for the densities $5 \times 10^{12} < n_e < 5 \times 10^{14}$ cm$^{-3}$ the 3p-3d excitation channel, which scales approximately as $n_e^2$, becomes comparable to the 3s-3d direct excitation from the ground state, making the ratio between the 3p and 3d level populations dependent on $n_e$. Moreover, due to the low excitation energies of these two levels the ratio between their populations is insensitive to the electron temperature for $T_e > 10$ eV. Also, these spectral lines could be observed in a single discharge due to their similar wavelengths, overcoming the shot-to-shot irreproducibility. Figure 3(a) shows the time-dependent population ratio for these two MgII levels. Also shown is the ratio predicted by the CR calculations using $n_e$ from Figure 3(b) and $T_e(t)$ determined in Sec. III.D. The error bars on $n_e$ are determined by the experimental error, the uncertainty in the electron temperature, and by the uncertainties in the atomic-physics modeling. The larger uncertainties at later times result from the lower electron temperature at these times. At t = 800 ns $n_e$ so determined matches $n_e$ determined from Stark broadening to within ±10%.

Figure 3(b) also shows a lower limit for $n_e$ obtained from the observed intensity of the ArI 6965-Å line. Here, we make use of the knowledge of the absolute density of the doped argon obtained from the ionization-probe array measurements [16]. By varying the electron temperature in the CR calculations as a free parameter we searched for the minimum electron density that would match the observed line intensity evolution, based on the known total ArI density. The presented lower limit accounts for the uncertainties in the ArI gas density and the absolute calibration of the spectroscopic system.

By determining $n_e$ at various positions in the x-z plane we constructed a 2D map of $n_e$ at various times. Such a map of the electron density in the x-z plane integrated over the y-direction at t = 1200 ns, which corresponds to the time of application of the POS high-current pulse, is shown in Figure 4. The density minimum at z = 3.5 cm is below the contact region of the two flashboards. The density peaks at z = 1.5 and z = 5.5 cm result from a faster plasma expansion at the ground-side edges of the flashboards. The lower density at the generator side (z = 1.5 cm) relative to the load side (z = 5.5 cm) is due to the lower current through the flashboard at the generator side, as a result of a larger inductance of the cables driving that flashboard.

The distribution of $n_e$ along the width of the electrodes (y-direction) is measured using observations along the z-axis, and the axially averaged electron density is observed to have a symmetric profile to within ±10% with a FWHM ≈ 12 cm. The density at the edges of the electrodes (along the y-direction) is found to be $(1-2) \times 10^{14}$ cm$^{-3}$ at t = 1200 ns.

## C.  Plasma Expansion velocity

In order to study the properties of the proton-plasma flowing at the front of the injected plasma, we observe light emission from the doped elements as they encounter the proton-plasma. The velocity of the proton plasma is determined from time-of-flight data obtained by observing the excitations of the dopants caused by the co-moving plasma electrons. The velocity of the slower carbon-dominated plasma is obtained from the time-of-flight data of spectral lines of carbon ions without doping.

Figure 5 shows the intensity of the doped HeI 6678-Å line at different x-positions. In these experiments, the POS electrodes are removed in order to avoid electrode-plasma effects on the measurement. The time delay between the helium injection and the application of the flashboard current is varied in the experiments in order to compensate for the different travel times of the gas to the various x-positions. The plasma front is seen to flow at a velocity of $(2.8\pm0.5)\times10^7$ cm/s, followed by a plasma with a higher density and a lower velocity of $(2.2\pm0.5)\times10^7$ cm/s, reaching the POS region at t = 400-500 ns. The slower rise of the helium signal at positions farther from the flashboard indicates that the velocity at which plasma leaves the vicinity of the flashboard drops in time.

The plasma that reaches the A-K gap 300-550 ns after the initiation of the flashboard discharge is primarily composed of protons with a small fraction of fast hydrogen. It is verified spectroscopically that the density of heavier ions such as carbon is negligible in this plasma front.

In order to determine the proton transverse velocity we observe the Doppler broadened $H_\alpha$ profile in experiments without doping and without the POS electrodes in order to only observe the hydrogen originating at the flashboard. Since only fast hydrogen atoms (those produced via proton charge-exchange near the flashboard surface) may reach the A-K gap so early, the $H_\alpha$ Doppler width should yield the proton transverse velocity.

Measurements of the $H_\alpha$ profile at x = 1 cm at early times (t = 400-700 ns) without gas injection and without POS electrodes show a constant Doppler profile with an average width that corresponds to a velocity of $(3.5\pm0.5)\times10^6$ cm/s in the y-direction. Comparing this velocity to the proton propagation velocity, obtained from time-of-flight, one can obtain the divergence angle of the plasma, found to vary from $\pm(7\pm1.5)^o$ at t = $320\pm40$ ns to $\pm(10\pm2)^o$ at t = $450\pm50$ ns and to $\pm(13\pm2.5)^o$ at t = $700\pm50$ ns.

Time-of-flight measurements of the CIII 2297-Å line without the POS electrodes showed a front propagating at a velocity of $(7\pm2)\times10^6$ cm/s, reaching the A-K gap at t ≈ 800 ns. This is followed by CIII with a higher density propagating at a velocity of $(5\pm1)\times10^6$ cm/s. The profile of this line indicated an average transverse velocity dropping from $(2.4\pm0.3)\times10^6$ cm/s for the front of the CIII plasma to $(1.7\pm0.2)\times10^6$ cm/s for the peak CIII density. The propagation velocity together with the transverse velocity yield a divergence angle of $\pm(20\pm8)^o$ for the CIII plasma.

### D. **Electron Temperature**

The electron temperature is determined as a function of time from the temporal evolution of absolute light intensities. To this end, the different effects that can affect the observed line intensities, such as variations of the ion or atom densities due to particle flow and ionization and the temporal variation of $n_e$, had to be considered. Here, by doping gaseous materials with injection velocities of the order of $10^5$ cm/s, we are able to minimize the effect of the flow. Using dopants with different ionization rates, it is also possible to study the effects of ionization.

The behavior of the HeI 1s3d $^1D$ level and the HeII n=4 level are modeled with the aid of CR calculations using the measured electron density (see Sec. III.B). Figure 6(a) shows the temporal evolution of these level populations, determined from the 6678-Å and the 4686-Å line intensities. The time-dependent electron temperature is determined by fitting the CR-code predictions for these level populations to the data. The modeling shown in Figure 6(a) is obtained using the electron temperature shown in Figure 6(b).

The modeling shows that the drop of the HeI 1s3d $^1$D population seen at t > 1 μs cannot result from ionization, due to the relatively low electron temperature at that time, indicating that this drop must occur due to a drop in $T_e$ as shown in Figure 6(b).

The electron temperature at t = 600-800 ns is also obtained from the absolute line intensity and knowledge of the density of the injected gas [16]. The absolute intensity of the HeI 6678-Å line shows that the electron temperature is 11±2 eV at t = 700 ns. Because the ionization of HeI is small, the obtained electron temperature time-dependence is unique, since the HeI 1s3d $^1$D level population is mainly affected by the variation of the excitation rates due to changes in the electron temperature ($n_e(t)$ is known).

A similar analysis is carried out for argon by measuring the temporal evolution of the ArI 6965-Å, ArII 4348-Å, and the ArIII 3286-Å lines. For ArI significant ionization occurs, requiring measurements of a few consecutive charge states in order to obtain a unique time dependence for the electron temperature. Figure 7(a) shows the temporal evolution of the ArI 4p'[1/2]$_1$, ArII 3p$^4$($^3$P)4p $^4$D$^o$, and the ArIII 3p$^3$($^4$S$^o$)4p $^5$P level populations, together with the predictions of the CR modeling. The experimental traces are averaged over 3 different discharges and have a temporal resolution of 20 ns. Figure 7(b) shows the electron temperature evolution that is used in the CR modeling to obtain the predictions in (a). The error bars on $T_e$ mainly result from the uncertainty in $n_e$.

The absolute intensity of the ArI 6965-Å line together with knowledge of the injected argon density show that $T_e$ = 14±2 eV at t = 550 ns. We also determined $T_e$ from the observed ionization time of CI produced from the disassociation of doped CH$_4$. We assume that the drop observed in the CI intensity at t > 900 ns results from ionization after the replenishing of neutral carbon through the various disassociation channels of CH$_4$ is finished. This assumption does not depend on a detailed knowledge of the chemistry of the CH$_4$ molecules and the resulting disassociation. This measurement is only performed at x = 1 cm, since in this position the line intensity of CI without the CH$_4$ doping is negligible. The ionization of CI at t = 900-1100 ns yields $T_e$ = 8±1.5 eV.

The density of the doped Ar, He, and CH$_4$ is varied from ≈2×10$^{13}$ to ≈10$^{14}$ cm$^{-3}$ in order to examine whether the gas doping caused cooling of the plasma electrons. For the low densities of doped He, the plasma electrons are found to cool by <<1 eV. However, for Ar and CH$_4$ doping the plasma electrons are found to cool by up to 2 eV (for a doped density of 10$^{14}$ cm$^{-3}$) resulting in an underestimate of $T_e$.

A summary of the electron temperature obtained using the temporal evolution of the helium and argon line intensities, the absolute intensities of helium and argon and the ionization time of CI, together with the error bars for x = 1 cm and z = 3.5 cm, is shown in Figure 8. $T_e$ is thus found to be 9±1.5 eV until t ≈ 1.0 μs, followed by a nearly linear drop to 5.0±0.5 eV at t ≥ 1.6 μs. Furthermore, the temperature at t = 1.2 μs (the time of the application of the POS current) is found to be $6.5^{+1}_{-0.5}$ eV near the cathode and somewhat lower near the anode ($T_e$ = $5.7^{+1}_{-0.5}$ eV). The agreement between the different methods and species substantiates the reliability of the determination of $T_e$.

Close to the generator-side edge of the plasma at z = -0.3 cm the electron temperature at t = 1.2 μs is found to be lower than at z = 3.5 cm, x = 1 cm. This seems to occur due to enhanced electron cooling caused by the solid anode surface at z < 0.

Also shown in Figure 8 is the electron temperature evolution determined using the temporal behavior of the HeI 1s3d $^1$D and the HeII n=4 spectral lines without the POS electrodes in place. It can be seen that the electron temperature without the POS electrodes in place is higher by ≈1 eV until t = 0.9 µs and by ≈2.5 eV at later times. The electron density at x = 1 cm is determined to be unaffected by the removal of the POS electrodes to within the accuracy of the measurement.

The determination of the electron temperature in present work was carried out assuming an equilibrium Maxwellian electron energy distribution. Nonetheless, it is known that in highly-transient plasmas of high-current discharges there may exist deviations from a Maxwellian EED [26], often in a form of an excess of high-energy electrons. The effect of the overpopulated tail of the electron energy distribution function, which is normally negligible for atomic processes with small energy thresholds ΔE (ΔE < $T_e$), can manifest itself especially in anomalous behavior of the highly-excited levels. This seems to be the case for the measured CIV lines, which originate from high lying levels (see Table 1). We were not able to consistently describe the absolute intensities of the CIV spectral lines with only a Maxwellian EED function, whereas an addition of a few percent of hot electrons (50-100 eV) resolved the discrepancy between modeling and measurements. We plan to address the issue of non-Maxwellian effects in the flashboard and POS plasmas in a separate publication.

Such a small deviation from a Maxwellian description of the plasma electrons does not devaluate the above-described quantitative determination of the electron temperature since it is obtained from emissions of low lying levels (11-23 eV) for which the contribution of the Maxwellian electrons overwhelms that of the deviation. Thus, $T_e$ serves as a measure of the mean electron energy in the flashboard plasma.

### E. Plasma composition

The plasma composition is determined from the observed light intensities and the known electron density and temperature. The total densities of HI, CI, CII, CIII, OII, and OIII are thus obtained from the intensities of the 6563, 2479, 2837, 2297, 4349, and 3047-Å lines, respectively. The effects of charge-exchange with neutral hydrogen on the observed lines of CII, CIII, OII, and OIII are estimated and found to be negligible.

This modeling in which the EED is assumed to be Maxwellian, yielded, however, unreasonably high densities for CIV. One possibility to explain these line intensities is to assume the presence of non-Maxwellian energetic electrons (as described in the previous section). It is estimated that a few percent of 50-100 eV electrons are enough for such an explanation. The other possibility is that the high CIV line intensities are caused by charge-exchange processes between CV and H resulting in selective population of some of the CIV levels.

The CV density could not be determined using line emissions from CV since no emission from CV lines was detected because the upper levels of all observable transitions are >300 eV above the ground state.

To examine the effect of charge-exchange on the CIV 5801 and 2530 Å lines we performed experiments with a hydrogen density varying from $3×10^{12}$ cm$^{-3}$ (by removing the POS electrodes and using no doping) to $1.5×10^{14}$ cm$^{-3}$ (keeping the POS electrodes in place and using H$_2$ doping) and observed the change in the CIV line intensities.

In the first configuration, in which the hydrogen density is too low for charge-exchange to contribute significantly to the level populations, are used to determine the density of CIV from CR calculations. It is important to note that since the CIV and CV originate at the flashboard (as described in Sec. III.F) their density is not affected by the removal of the POS electrodes.

In the experiments with $H_2$ doping the relative velocity between the hydrogen and CV, required for the charge-exchange modeling, is obtained from the flow velocity of CV, found to be $(5\pm1)\times10^6$ cm/s (the doped hydrogen is slow). The hydrogen density determined from the $H_\alpha$ intensity and the state-selective cross sections for the charge exchange [27,28] are then used to determine the density of CV.

The large uncertainty in the CV density mainly results from the uncertainty in the impact velocity and the resulting error of the state-selective charge exchange into the 3p level, which drops very sharply at low impact velocities, as the dominant channel for charge exchange becomes the 3d level [27] (the charge exchange into the 3d level has nearly no effect on the 3p population since the 3d level radiatively decays to the 2p level). A second uncertainty in the CV density results from the uncertainty in the $H_2$ density. The cross section for charge exchange with these molecules is higher than the cross section with hydrogen atoms, especially at low velocities where the 3p level remains the dominant channel of charge exchange with $H_2$ [29].

A lower limit for the CIV density can be obtained from the level populations observed during the POS operation when the electron temperature is found to rise [30] and the CIV level populations become dominated by electron excitation rather than by charge exchange processes. By taking the optimum excitation rates in the CR calculations, this analysis results in a lower limit for the CIV density, giving for $t = 1.2$ μs $(6\pm1.5)\times10^{13}$ and $(2\pm0.5)\times10^{13}$ cm$^{-3}$ at $x = 2.2$ and $x = 1$ cm, respectively. An upper bound for the CIV density is the measured electron density and the requirement for charge neutrality.

Measurements also show that the density of other heavier elements such as iron, copper, and aluminum are negligibly small except at a distance less than 1 mm from the electrode surfaces. This results from the low flow velocity of these ions into the POS region. Near the POS electrodes, however, heavy impurities, desorbed from the surfaces due to the impact of the primary flashboard plasma, are found to make up a few percent of the plasma density.

The determination of the densities of most of the plasma constituents allows the proton density to be obtained from the balance between $n_e$ and the total non-protonic ionic charge. However, due to the large uncertainty in the density of CIV and CV there is a significant uncertainty in the proton density.

Table II summarizes the plasma composition at $x = 1$ (the middle of the A-K gap) and $x = 2.2$ cm (near the anode) for $z = 3.5$ cm (the axial center of the POS) integrated along the y-direction. The composition is for the time 1.2 μs after the start of the flashboard current pulse.

## F. **Origin of the various plasma constituents**

Measurements with and without the electrodes are used to distinguish between plasmas produced by the flashboard and secondary plasmas originating from the POS electrodes. The electron density and temperature, required for the CR modeling of the observed line intensities, are determined for both configurations. The electron density is found to be approximately the same while the electron temperature is found to be higher at t > 0.9 µs if the POS electrodes are removed (see Figure 8).

In experiments with the POS electrodes, the observed intensity of $H_\alpha$ is found to increase by more than an order of magnitude as a result of hydrogen originating from the electrode surfaces. Comparison of the hydrogen density near the anode and cathode shows that the hydrogen density emanating from the anode is approximately two times higher than that from the cathode. In experiments with the POS electrodes (without doping) the Doppler-dominated $H_\alpha$ width at t = 600-900 ns is found to depend on the location of the observation. In the middle of the A-K gap (x = 1 cm) $H_\alpha$ has a broad profile corresponding to an average hydrogen velocity of $(4.5\pm0.3)\times10^6$ cm/s in the y-direction. Experiments near the anode (x = 2 cm) show a narrow profile with an average velocity in the y-direction of only $(2.5\pm0.3)\times10^6$ cm/s. This data can be interpreted by assuming an injection of fast and slow hydrogen from the electrodes with a large divergence angle ($\pm45^o$). As a result, the fast hydrogen reaches the middle of the A-K gap at t = 600 – 900 ns, whereas the denser and slower hydrogen only reaches the x = 2 cm position and the x=0.5 cm for hydrogen ejected from the cathode.

The density of CII and CIII at x = 2 cm is found to be approximately the same for the two configurations until t ≈ 1 µs, after which the density of these ions increases sharply in the configuration with the electrodes and becomes about 6 times higher than without electrodes at t = 1.2 µs. A similar increase is also seen near the cathode (x = 0.3 cm). This means that most of the high density carbon plasma near the anode and cathode originates from the electrode wires rather than from the flashboard. However, unlike CII, CIII, and hydrogen, we may conclude that CIV and CV originate at the flashboard, since the temperature in the anode-cathode gap is too low to ionize CIII originating from the electrodes within ≈1 µs.

## IV. **Summary and Discussion**

Plasma doping with various gaseous materials allows for studying many properties of the flashboard-generated plasma. Both Stark broadening of hydrogen lines and spectral line ratios of MgII are used to determine the evolution of $n_e$. The agreement between the two methods at t = 800 ns is within ±10%. The measurement of spectral line ratios of MgII also allows for accurate determination of the temporal evolution of $n_e$ for electron densities as low as $1\times10^{13}$ cm$^{-3}$. At the time at which the POS is operated $n_e$ is found to vary between $3\times10^{14}$ cm$^{-3}$ near the cathode to $7\times10^{14}$ cm$^{-3}$ near the anode.

The electron temperature of the expanding flashboard plasma at x = 1 cm, determined from the temporal evolution of different doped-species spectral lines, is found to drop from 18±3 eV at t = 500 ns to 10.5±1.5 eV at t = 800 ns and to $6.5^{+1}_{-0.5}$ eV at t = 1.2 µs. In experiments in which the POS electrodes are removed, the drop of the electron temperature is slower and an electron temperature of 10±1 eV is found at x = 1 cm at t = 1.2 µs (see Figure 8). The difference between the two configurations is explained by the cooling of the plasma due to the interaction with the electrodes.

As mentioned in Sec. III.D, the EED in our plasma seems to deviate from a Maxwellian one, with an excess of high-energy (~ 50-100 eV) electrons. We estimate that electrons with an energy of 100 eV have a mean free path of 1-3 cm for electron-electron collisions for an electron density of $(1-3)\times 10^{15}$ cm$^{-3}$ (this density is taken as the average electron density in the region between the flashboard and the POS, i.e., x > 2.6 cm). Thus, high-energy electrons produced near the flashboard may reach the POS region without thermalization, giving rise to the deviation of the EED from a Maxwellian.

The relatively high accuracy in the determination of $n_e$ and $T_e$ allows for a reasonably reliable determination of the plasma composition. To this end, the effects of charge exchange processes on the line intensities are accounted for. It is shown that the possible presence of non-thermal electrons with energies above 50 eV mainly affects the uncertainty in the determination of the CIV density. In the center of the A-K gap the plasma is shown to consist of CIV with lower densities of CII, CIII, and CV. Hydrogen is found to be the only neutral element with a significant density (<10% of $n_e$). The proton density determined from the balance between $n_e$ and the total non-protonic ionic charge has a large uncertainty due to the relatively large uncertainty in the CIV and CV densities. The density of oxygen ions is 3 - 10 times lower than that of carbon ions and the density of heavier ions (such as SiII, FeII, and NiII) are estimated to be less than $10^{12}$ cm$^{-3}$. This work demonstrates the need for detailed CR modeling in order to avoid large errors in the determination of the electron temperature and plasma composition.

By combining the electron density measurements with time-of-flight measurements and with the knowledge of the plasma composition it is possible to study the velocity and origin of the various plasma constituents. Here, we make the distinction between four different plasmas. The first plasma consists of protons traveling at a velocity up to $(2.8\pm 0.3)\times 10^7$ cm/s and with a density less than $1\times 10^{13}$ cm$^{-3}$. This is followed by a slower and more dense proton plasma with a propagation velocity of $(2.1\pm 0.3)\times 10^7$ cm/s and a density of $2\pm 1\times 10^{13}$ cm$^{-3}$. The protons in both plasmas are found to have a very low divergence angle $\pm(7-13)^o$. The third plasma is found to mainly consist of carbon ions and protons with smaller concentrations of hydrogen, oxygen and silicon ions. The electron density of this plasma increases from $10^{14}$ cm$^{-3}$ to >$10^{15}$ cm$^{-3}$ in the A-K gap. The most abundant charge state of the carbon ions in this plasma is CIV.

The fourth plasma originates at the POS electrodes as a result of the bombardment by the proton plasmas. This plasma mainly consists of hydrogen, CII, CIII, and smaller contributions of OII and OIII and contributes most of these species in the A-K gap as is found by comparison between experiments with and without the POS electrodes. The carbon ions originating at the electrodes have an expansion velocity of $(2\pm 0.5)\times 10^6$ cm/s, thus they are confined to a region < 1 cm from the electrodes at the time the POS is operated. This is also correct for most of the hydrogen ejected from the electrodes that has an expansion velocity of $(1.5\pm 0.5)\times 10^6$ cm/s. However, approximately 20% of the hydrogen travels at a higher expansion velocity, $(3-5)\times 10^6$ cm/s, thus filling the entire A-K gap at the time the POS is operated.

In our configuration, the effect of plasma ejection from the electrodes on $n_e$ is small since the electrodes consist of wires with a 94% transparency and the time delay between the creation of the electrode plasmas and the POS operation is small. This is also the reason why in our configuration the most abundant carbon species is CIV, which originates at the flashboard. In another configuration used earlier, in which a solid cathode was used (rather than highly transparent) and the time delay between the flashboard discharge and the POS operation was longer, the density of the secondary electrode plasma in the A-K gap was higher than that of the flashboard plasma. One may, therefore, suggest that in other experiments [31,32], in which low-transparency POS electrodes are used and the time delay between the flashboard discharge and the POS operation is longer the effect of electrode plasmas is more important. This may influence the plasma composition (more ions with lower charge states), the electron temperature (lower $T_e$) and electron density (higher $n_e$) in most of the A-K.

These results demonstrate that the properties of the prefilled plasma used in pulsed-power experiments, besides being influenced by the plasma-source geometry and the discharge circuit, can be highly affected by material release from the electrodes. This effect probably depends on the flux of plasma that strikes the electrode surfaces, on the electrode-surface area and conditions, and on the time delay between the source discharge and the system operation.

Generally, knowledge of the plasma properties, and in particular the plasma composition, is highly important for understanding the device operation. We believe that the diagnostic methods here described can be used to systematically study most of the important plasma properties in a variety of experimental conditions and geometries.

## ACKNOWLEDGMENTS

The authors are grateful to Y. Krasik and A. Weingarten for highly illuminating discussions, to V. A. Bernstam for aiding with the CR modeling, to I. Bray for providing us with atomic data and to P. Meiri for his technical assistance. This work is supported by the Minerva foundation (Munich, Germany).

**Tables**:

Table I: A list of the spectral lines used for the determination of $n_e$, $T_e$, and the composition of the flashboard plasma.

| Species | Wavelength | Upper level of | Energy of the |
|---|---|---|---|
| **Plasma constituents** | | | |
| H | 6563 | n=3 | 12.1 |
| H | 4861 | n=4 | 12.7 |
| CI | 2479 | 2p3s $^1P^o$ | 7.7 |
| CII | 2837 | 3p $^2P^o$ | 16.3 |
| CIII | 2297 | 2p$^2$ $^1D$ | 18.1 |
| CIII | 4647 | 2s3p $^3P^o$ | 32.3 |
| CIV | 5802 | 3p $^2P^o$ | 39.7 |
| CIV | 2530 | 5g $^2G$ | 55.8 |
| CV | 2271 | 1s2p $^3P^o$ | 304.4 |
| OII | 4349 | 2p$^2$($^3P$) 3p $^4P^o$ | 25.8 |
| OIII | 3047 | 2p3p $^3P$ | 37.3 |
| OIV | 3063 | 3p $^2P^o$ | 48.3 |
| SiIII | 2542 | 3p$^2$ $^1D$ | 15.2 |
| CuII | 2545/2403 | 5s $^3D$ | 13.4 |
| **Doped elements** | | | |
| HeI | 5875 | 1s3d $^3D$ | 23.1 |
| HeI | 6678 | 1s3d $^1D$ | 23.1 |
| HeII | 4686 | N=4 | 51.0 |
| MgII | 2795.5 | 3p $^2P^o$ | 4.5 |
| MgII | 2798 | 3d $^2D$ | 8.9 |
| ArI | 6965 | 4p'[1/2]$_1$ | 11.5 |
| ArII | 4348 | 3p$^4$($^3P$) 4p $^4D^o$ | 19.5 |
| ArIII | 3286 | 3p$^3$($^4S^o$) 4p $^5P$ | 25.4 |

Table II: The plasma composition at z = 3.5 cm 1.2 μs after the start of the flashboard current.

| Species | Plasma composition at x=1.0 cm [cm$^{-3}$] | Plasma composition at x=2.2 cm [cm$^{-3}$] |
|---|---|---|
| H | $(1.5\pm0.5)\times10^{13}$ | $(8\pm3)\times10^{13}$ |
| CI | $(1.1\pm0.3)\times10^{12}$ | $(2\pm0.7)\times10^{13}$ |
| CII | $(1.3\pm0.4)\times10^{13}$ | $(6\pm2)\times10^{13}$ |
| CIII | $(2.2\pm0.8)\times10^{13}$ | $(9\pm3)\times10^{13}$ |
| CIV | $(7\pm4.5)\times10^{13}$ | $(1\pm0.4)\times10^{14}$ |
| CV | $(2\pm1)\times10^{13}$ | $(2\pm1)\times10^{13}$ |
| OII | $(1\pm0.4)\times10^{12}$ | $(1\pm0.4)\times10^{13}$ |
| OIII | $(4\pm2)\times10^{12}$ | $(2\pm1)\times10^{13}$ |
| OIV | $(1\pm0.6)\times10^{13}$ | $(3\pm2)\times10^{13}$ |
| $\Sigma n_i z_i$ (not Protons) | $3.9\times10^{14}$ | $7.5\times10^{14}$ |
| Protons from $n_e - \Sigma n_i z_i$ | $(0\text{-}2.5)\times10^{14}$ | $\approx 0\times10^{14}$ |
| $n_e$ | $(4.5\pm1)\times10^{14}$ | $(7\pm1)\times10^{14}$ |

# Figure Captions:

Figure 1: Schematic of the POS, the flashboard plasma sources, the gas-doping arrangements, and the spectroscopic system. The inter-electrode region of the planar POS is prefilled with plasma from two flashboard plasma sources. A fast gas valve, a nozzle, and a skimmer are used to locally dope the plasma with various species. Lenses and mirrors are used to collect light from the doped column into the spectrometer. A cylindrical lens focuses the output of the spectrometer onto a fiber bundle array that passes the light to 12 photomultiplier tubes.

Figure 2: The time dependent electron density of the flashboard plasma for $z = 3.5$ cm (the middle of the plasma in the z-direction), determined from Stark broadening of hydrogen lines. Two traces obtained in different discharges are shown for $x = 1$ (near the middle of the A-K gap) and $x = 2$ cm (near the anode).

Figure 3: (a) The observed time-dependent ratio of the MgII 3p $^2P^o$ and 3d $^2D$ level populations time-averaged using a 30 ns filter and averaged over three experiments. The measurement is at $x = 1$, $y = 0$, and $z = 3.5$ cm (the middle of the plasma in the z-direction). The oscillations visible prior to $t < 500$ ns are caused by the weak signal of the MgII 2798 Å line and the resulting poor photon statistics. Also shown is the level population ratio predicted by the collisional-radiative modeling. (b) The electron density used in the CR calculations to obtain the predicted ratio shown in (a). Also shown is $n_e$ obtained from Stark broadening at $t \geq 780$ ns. The thick solid curve represents a lower limit for $n_e$ obtained from the absolute intensity of the ArI 6965-Å line.

Figure 4: A two dimensional map of $n_e$ integrated over the y-direction 1.2 μs after the operation of the flashboard. The spatial resolution in these measurements is ±0.5 cm in the z-direction and ±0.2 cm in the x-direction, respectively. The errors are $\pm 0.7 \times 10^{14}$ cm$^{-3}$.

Figure 5: Time-of-flight measurements of the proton-plasma obtained from the HeI 6678-Å line intensity at various distances from the flashboard surface in the middle of the plasma in the z-direction. The results are averaged along 8 cm in the y-direction. a(b) cm/s means $a \times 10^b$ cm/s. Note that larger x-positions are closer to the flashboards.

Figure 6: (a) The observed temporal evolution of the HeI 1s3d $^1D$ and the HeII n=4 level populations (symbols) at $x = 1$ and $z = 3.5$ cm. Two traces corresponding to different experiments are shown for HeI. Also shown (thin lines) are the best-fit level populations predicted by the collisional-radiative modeling using the electron temperature evolution shown in (b).

Figure 7: (a) The observed temporal evolution of the ArI 4p'[1/2]$_1$, the ArII 3p$^4$($^3$P)4p $^4D^o$, and the ArIII 3p$^3$($^4S^o$) 4p $^5P$ level populations (symbols). The results are averaged over 3 cm along the y-direction, at $x = 1$ and $z = 3.5$ cm. Also shown (thin lines) are the level populations predicted by the CR modeling. (b) The electron temperature evolution used in the CR code to obtain the curves shown in (a).

Figure 8: The temporal evolution of the electron temperature at x = 1, y = 0, z = 3.5 cm obtained using the various methods discussed in the text. Also shown is the evolution of $T_e$ with no POS electrodes.